\documentclass[useAMS,usenatbib,usegraphicx]{mn2e}

\title[Mass and radius of the neutron star in 4U 1820-30 ]{ Mass and 
  radius estimation for the neutron star in X-ray burster 4U 1820-30 }

\author[K. Ku\'smierek, J. Madej and E. Kuulkers]{
K. Ku\'smierek$^{1}$\thanks{E-mail: krzysztof.kusmierek@gmail.com (KK);
$\hspace{20mm}$ jm@astrouw.edu.pl (JM); Erik.Kuulkers@esa.int (EK) },
J. Madej$^{1}$, E. Kuulkers$^{2}$ \footnotemark[1] \\
$^{1}$ Warsaw University Observatory, Al. Ujazdowskie 4,
       00-478 Warsaw, Poland  \\
$^{2}$ ISOC, ESA, European Space Astronomy Centre (ESAC), P.O.~Box 78, 28691, 
       Villanueva de la Ca\~nada (Madrid), Spain    }

\begin{document}
\date{Accepted .................................. }
\pagerange{\pageref{firstpage}--\pageref{lastpage}} \pubyear{2011}

\maketitle
\label{firstpage}

\def\te{T_{\rm eff}}

\begin{abstract}
We present a new method for determining masses and radii of neutron 
stars residing in thermo-nuclear X-ray burst sources. To illustrate this 
method we apply it to a burst from the source 4U 1820-30 recorded
by the {\it Rossi X-Ray Timing Explorer}.
Fits of the observed X-ray spectra to grids of Comptonised model 
atmospheres yield estimates for the mass and radius of the neutron star,
$M=1.3 \pm 0.6M_\odot$ and $R=11^{+3}_{-2}$ km, respectively.
\end{abstract}


\section{Introduction}

The low-mass X-ray binary (LMXB) 4U 1820-30 is located in the core of the
globular cluster NGC 6624 (Giacconi et al. 1974). Using optical observations
the distance is estimated to be $7.6 \pm 0.4$ kpc (Kuulkers et al. 2003).
Thermo-nuclear X-ray bursts (so-called type~I X-ray bursts, see, e.g., 
Lewin, van Paradijs \& Taam 1993, Strohmayer \& Bildsten 2006, for reviews)
from this source were discovered by Grindlay et al. (1976). 
4U 1820-30 has an extremely short binary orbital period of 685 sec 
(Stella, Priedhorsky \& White 1987). Such a short orbital period and the
X-ray burst activity imply that it is a low-mass binary system with a 
$\sim 0.06-0.07 M_\odot$ secondary (donor) star orbiting a neutron star. The
donor is most probably a helium white dwarf (Rappaport et al. 1987).

Almost all of the X-ray bursts from 4U 1820-30 release so much energy, 
that the surface luminosity of the neutron star reaches the Eddington limit
(see Vacca et al. 1986, Haberl et al. 1987, Kuulkers et al. 2003, Galloway
et al. 2008). The atmosphere then expands due to radiation pressure and the
photospheric radius increases by a factor of up to $\sim 20$.
At the time of expansion and subsequent contraction the X-ray luminosity of
4U 1820-30 is expected to remain almost constant near the Eddington limit.
Such a scenario implies that at first the effective temperature $\te$ of
the compact star decreases when the radius increases. At the subsequent
stage, the radius decreases and the $\te$ increases.

X-ray spectra taken during such strong burst can be used to estimate mass
and radius for the compact star in 4U 1820-30, and hence, for determining 
the equation of state (EOS) for ultradense matter hidden in the stellar 
interior (cf. Haberl et al. 1987; Lewin et al. 2003).

Measurements of masses and radii of neutron stars are necessary for
better understanding of the populations of all stellar objects and the
latest stages of stellar evolution. Moreover, it is possible that some of
them are made out of strange matter with the equation of state (EOS) different
than that for neutrons. In such a case mass $M$ and radius $R$ of the
compact star can be apparently lower than for a neutron star (see e.g. 
Haensel et al. 2007). Mass and radius determination for a compact star is
understood as an intermediate step to constrain EOS in many recent papers
(\"Ozel 2006; Lattimer \& Prakash 2007 for example).

At present, masses and radii of weakly magnetized compact stars in
X-ray burst sources are rather poorly constrained, due to several reasons.
First, usually there exists several sources of X-ray emission contributing
to the total flux: thermonuclear explosions in the stellar envelope,
accretion disk and accretion stream, and the transition layer between
disk and the stellar surface. Second, observed satellite X-ray spectra are
of poor quality and their interpretation with spectral fitting software
is ambiguous. Third, available models of X-ray emission (and model atmospheres
of hot compact stars) mostly are schematic and inadequate
for fitting of non-Planckian spectra.
The advantage of this work is that we use much more sophisticated model
spectra than those used in previous papers 
(see the last paragraph of Section 2 and the beginning of Section 4).

\section{Outline of the method}

Our analysis uses published {\it Rossi X-ray Timing Explorer} ({\sc RXTE}) 
Proportional Counter Array (PCA) X-ray spectral data of this source 
(Kuulkers et al. 2003), which were obtained during a strong burst just after
the phase corresponding to the maximum effective temperature $\te$. 
We assume, that the radius of the neutron star photosphere had decreased
to its normal value. The cooling phase had just started, but the bolometric
(X-ray) luminosity is still very close to the Eddington luminosity.  

At first, we determined the surface gravity $\log g$ and the surface 
gravitational redshift $z$ for 4U 1820-30. Both parameters were constrained
by the best fits of the observed spectra to grids of hundreds of theoretical
X-ray spectra of hot neutron stars. Finally, the mass $M$ and radius $R$
of the compact object were determined from simple algebraic relations 
corresponding to a non-rotating star and the Schwarzschild metric 
(Majczyna \& Madej 2005).

Rotational distortion of a compact star depends both on the rate of rotation
and details of its internal structure including the EOS. None of them are
known a priori. Also there are no available models of flattened compact 
stars. Therefore, assumption of a nonrotating star is the only practical
approximation for our research.

We also assume that the burst emission from the source is isotropic.
The assumption means that the eruption exhibits spherical symmetry, and
anisotropy caused by the disk was neglected. 
Moreover, properties of the
accretion disk (e.g. its thickness), possible existence of the accretion
disk corona and the unknown inclination angle of the disk do not alter
the observed X-ray spectrum by assumption.

Grids of helium-dominated model atmospheres and X-ray spectra of hot neutron
star were computed with the {\sc ATM21} code, which computes model atmospheres
with the account of Compton scattering on free electrons. The code takes
into account angle-averaged Compton scattering of X-ray photons with initial
energies approaching the electron rest mass.
Our models atmospheres were computed with rich set of bound-free and free-free
energy-dependent opacities. A detailed description of the
equations and numerical methods were given in a long series of earlier papers
(Madej 1991a,b; Joss and Madej 2001; Madej, Joss and R{\'o}{\.z}a{\'n}ska
2004; Majczyna et al. 2002, 2005).

\section{Observational data}

Our analysis makes use of the series of 113 X-ray spectra taken during 
a strong burst from 4U 1820-30 recorded by the PCA onboard the {\it RXTE}
satellite on UT 1997 May 2, which started at 17:32:45.
The burst spectra are at a 0.25 sec time resolution, and are numbered
from 5 to 120. They are corrected for deadtime.
All PCUs were on during the observation.
We subtracted the pre-burst persistent emission from the burst emission.
The whole sequence of spectra was described and analyzed in detail in the
paper by Kuulkers et al. (2003). For this paper we took into account the
latest response matrix, using {\sc marfrmf} version 3.2.6.

For our analysis we have chosen only 2 spectra of that series, which 
correspond to the phase of the burst just after the maximum
blackbody temperature $T_{\rm bb}$, see Kuulkers et al. (2003).  
These spectra are labeled No. 21 and 22. The single burst of 4U 1820-30
was chosen as trial and validation of the method. 

\section{Models}
\label{sec:models}

Model atmospheres and theoretical spectra of our grid were parametrized by
\begin{description}
\item[--] effective temperature $T_{\rm eff}$ in the range $1.0 \times 10^7
 < T_{\rm eff} < 3.0 \times 10^7$ K in steps of $0.2 \times 10^6$, and
\item[--] decimal logarithm of gravitational acceleration in the atmosphere
of a neutron star, $\log g$. The lowest gravity model of $\log g_{\rm min}$
just touches the critical gravity, when the atmosphere loses 
hydrostatic equilibrium (models at $\log g$ equal to $\log g_{\rm min}-0.1$
or less were dynamically unstable). The highest gravity equals to 
$\log g_{max}=15.0$ for all effective temperatures $T_{\rm eff}$. 
E.g. for $T_{\rm eff} = 2.6 \times 10^7$ K we precomputed a set of
theoretical spectra corresponding to $\log g = 14.3$, $14.4$, $14.5$, $14.6$,
$14.7$, $14.8$, $14.9$ and $15.0$. Our grid consists of 288 models.

\end{description}

We assumed for all analyzed spectra, that the photospheric radius of the
compact object has already decreased to its normal value. The cooling phase
has just started, but
the bolometric luminosity is still close to the Eddington luminosity.
In this regime model atmospheres and their X-ray spectra are more
sensitive for changes of both the effective temperature $T_{\rm eff}$
and gravity $\log g$. The farther off the Eddington luminosity, the less
accurate is our method of determining both parameters of the compact star.
In practice, it turned out that we can use it only for the spectra No. 21 and 22.
To make sure that our minimum of $\chi^2$ is a global,
we have used an extensive grid of models.

For a given $T_{\rm eff}$ calculated X-ray spectra exhibit peak fluxes always
at an energy higher than the blackbody of the same temperature. It is a 
well-known effect that the $T_{\rm bb}$ from black-body fits to the burst
spectra (called color temperature $T_{col}$) are always greater than $\te$.
In other words, color factor $f=T_{col}/T_{bb} > 1$ (Madej 1974; Lewin et
al. 1993; Shaposhnikov \& Titarchuk 2004).
We draw attention of the reader to the fact that this
difference is implicitly taken into account in our calculated burst spectra.

The important issue is the choice of the chemical composition of the model 
atmospheres. We know that the material accreted by the neutron star
is dominated by helium with small amounts of hydrogen (Cumming 2003). Moreover, 
at the phase of a burst just after the maximum luminosity some ashes from
nuclear burning are still exposed at the neutron star surface. That matter
was ejected by the radiation driven wind during the phase of radius expansion
(Weinberg, Bildsten \& Schatz 2006). However, we accepted a rather simple
chemical composition for our grids of model atmospheres, i.e., we assumed
helium with small amounts of iron. In this way iron was used as the ''average
metal'' and replaced a mixture of heavy elements which must be created during
a thermonuclear burst.

We were able to constrain the iron abundance by estimating the quality of 
the trial fits, which was given by the value of $\chi^2$ for 1 degree of
freedom.We tested fits to the grid of almost pure helium atmosphere spectra
with very small amount of hydrogen and to few grids of model 
atmospheres adding iron of various abundance. (Hydrogen must be included
in the {\sc ATM21} input data set due to numerical reasons.)
For the final fitting we used the grid of model atmospheres and spectra 
computed for the best chemical composition, as shown in Table 1.

\section{Fitting procedure}
\label{sec:fit}

All fits of the observed X-ray spectra to the numerical models depend
on the following parameters:

\begin{description}
\item[--] hydrogen column density $N_H$. We assumed that $N_H$ is 
   identical for all spectra and equals 
   to $N_H = 0.16\times 10^{22} \pm 0.003$ cm$^{-2}$
   (see Kuulkers et al. 2003),
\item[--] redshift $z$, from $0.1$ to $0.6$, and
\item[--] normalization parameter.
\end{description}

\begin{table}
\caption {Number abundances of elements, $N_Z=N_{el}/N_H$ }
\hspace{1.0cm}
\begin{tabular}{||c|c||}
\hline \hline
      $Z$       & $N_Z$   \\ \hline

       1        & 1.00e+00   \\
       2        & 1.00e+06   \\
      26        & 1.11e+03   \\
\hline \hline
\end{tabular}
\label{tab:chemical}
\end{table}

For each of the combination ($\te$, $\log g$) of the available 
grid of numerical models of X-ray spectra we determined the redshift $z$,
the normalization parameter and $\chi^2$. $\chi^2$ was determined 
from the fits using the {\sc XSPEC} software, version 11.3.2.
Within {\sc XSPEC}, we used the spectral fitting model 'wabs atable' 
in the energy range $2.9-20.0$ keV.

\begin{itemize}
\item[{\bf 1.}] We ignored models with limiting redshift $z=0.1$ and $z=0.6$.
Remaining values of $\log g$ and the corresponding fits were rejected.
\item[{\bf 2.}] We have chosen only models with $\chi^2$ in the range
$[\chi^2_{min},\chi^2_{min}+\Delta\chi^2]$, where $\chi^2_{min}$ is the minimum
value $\chi^2$ and $\Delta\chi^2$ represents the increase of $\chi^2$ small
enough to be in the $1\sigma$ confidence range. We applied here the estimate
for $\Delta\chi^2$ taken from Press et al. (1996),
page 692, which give in our case $\Delta\chi^2=2.3/23$ d.o.f$=0.1$.
For the spectrum No. 21 we obtained $\chi^2_{min}=1.257$ (see Fig. 1) and
for the spectrum No. 22 we obtained $\chi^2_{min}=0.816$ (see Fig. 2).

\begin{figure}
\resizebox{1.05\hsize}{!}{\rotatebox{270}{\includegraphics{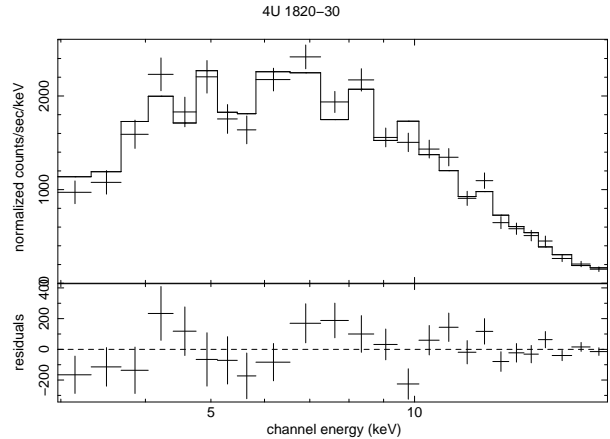}}}
\caption{ Sample fit of our redshifted model spectrum of $T_{\rm eff} = 
  2.5\times 10^7$ K, $\log g = 14.2$ and the redshift $ z=0.24$ to the RXTE
  spectrum No. 21 of this X-ray burst. }
\label{fig1}
\end{figure}

\begin{figure}
\resizebox{1.05\hsize}{!}{\includegraphics[width=\hsize,angle=270]{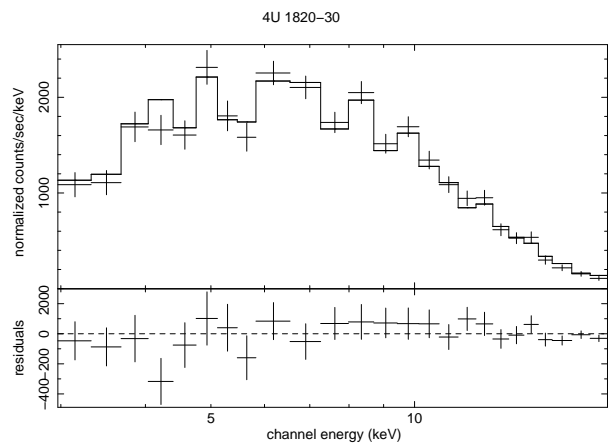}}
\caption{ Sample fit of the spectrum No. 22 of 4U 1820-30 by our redshifted
  model with $T_{\rm eff} = 2.2\times 10^7$ K, $\log g = 14.0$ and $z=0.12$. }
\label{fig2}
\end{figure}

\item[{\bf 3.}] For each model  which has passed the above procedure
we obtained a single pair of values for mass $M$ and
radius $R$ from the algebraic equations corresponding to nonrotating 
star and the Schwarzschild metric, cf. Majczyna \& Madej (2005).
The gravitational redshift is given by:
\begin{equation}
   1+z = \left(1-\frac{2GM}{Rc^2}\right)^{-1/2}
\end{equation}
where $G$ is the gravitational constant, $M$ is the neutron star mass,
$R$ is the radius measured on the neutron star surface, $c$ denotes 
the speed of light.
Gravitational acceleration on the neutron star surface equals to :
\begin{equation}
g=\frac{GM}{R^2}\left(1-\frac{2GM}{Rc^2}\right)^{-1/2}
\end{equation}
We solve Eqs. 1-2 for mass $M$ and radius $R$, and obtain the
following explicit expressions:
\begin{equation}
\label{eqR}
R=\frac{z\, c^2}{2\, g}\,\frac{(2+z)}{(1+z)}
\end{equation}
\begin{equation}
\label{eqM}
M=\frac{z^2 c^4}{4gG}\, \frac{(2+z)^2}{(1+z)\, ^3}
\end{equation}

Both mass and radius of a neutron star are functions only of the surface
gravity $g$ and the gravitational redshift $z$. The effective temperature
$T_{\rm eff}$ does not directly influence neither $M$ nor $R$.

We assumed here that for each fitted X-ray spectrum the apparent radius
of the photosphere $R_{\rm ph}$ at the touchdown phase of that burst is
equal to the true radius $R$ of the neutron star in 1820-30. This is a widely
accepted assumption, that both radii are equal at the touchdown phase of
strong X-ray bursts with radius expansion. Note, that it was critically
rediscussed recently by Steiner et al. ( 2010).

Extreme masses and radii collected in the above set determine the errors
of mass and radius of the $1\sigma$ confidence range.
\item[{\bf 4.}]
The best values of mass $M$ and radius $R$ are arithmetic averages of
individual determinations. We decided to use the mean value rather than the
determination based on the model with $\chi^2_{min}$, since our grid of models
is extensive but was not fine enough.
\end{itemize}

For the spectrum No. 21 the we obtained
mass $M=1.19^{+1.28}_{-0.75}M_\odot$ and radius 
$R=9.40^{+8.91}_{-3.40}$ km. For the spectrum No. 22, 
$M=1.36^{+0.75}_{-0.71}M_\odot$ and $R=11.38^{+2.83}_{-2.23}$ km, 
respectively. Both determinations yield similar values, but the spectrum
No. 22 yields much lower errors.

We had to average measurements with asymmetric errors. This difficult problem
was analysed by Roger Barlow (arXiv:physics/0406120v1 and 
arXiv:physics/0401042v1).
We used his Java applet which is available at
http://www.slac.stanford.edu/{$\sim$}barlow/statistics.html
and obtained averaged results: $M=1.3 \pm 0.6M_\odot$,
$R=11^{+3}_{-2}$ km (see Fig. 3).

\section{Conclusions}

In this paper we presented a new method to determine the mass and radius 
of the compact object in type~I X-ray burst sources. We analyzed X-ray spectra
of the burster 4U 1820-30 in phases when the luminosity of the source is
still close to the Eddington luminosity and the compact star photosphere
returned close to its normal size after radius expansion.
The observed X-ray spectra were fitted to grids of theoretical spectra
of model atmospheres in radiative and hydrostatic equilibrium computed with
the precise treatment of Compton scattering on free electrons.

Blackbody fits allow us to determine radius only for the burster of known
distance. Using our method we may obtain not only radius but also mass of
the neutron star simultaneously. Furthermore, our $M$ and $R$ determinations
are not directly dependent on the distance to the source. It is affected
only by the value of the hydrogen column density $N_H$.
This is a great advantage over blackbody fits.

\begin{figure}
\resizebox{\hsize}{0.8\hsize}{\includegraphics{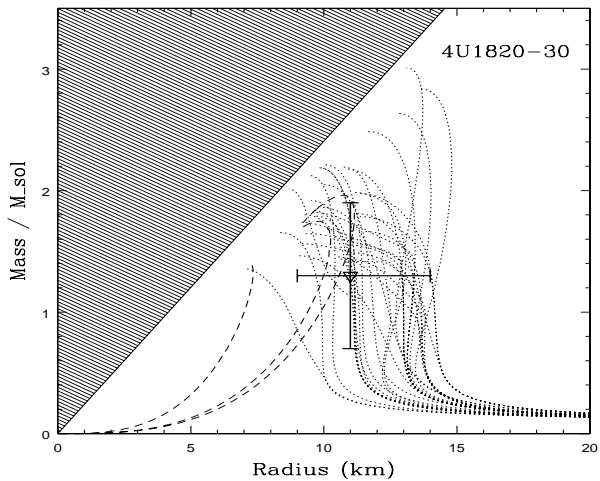}}
\caption[]{Mass and radius for the compact star in 4U 1820-30. 
Triangle represents the average values derived from spectra No. 21 and 
No. 22. We adopted Fig. 2 from Bejger \&
Haensel (2002), who plotted a large sample of equations of state for neutron
stars and strange stars (solid and short-dashed lines, respectively). 
Shaded area corresponds to a exclusion zone in parameter space from 
General Relativity combined with the constraint $v_{\rm sound} <  c$. }
\label{fig:fig3}
\end{figure}

Using our method, we constrained the mass 
$M=1.3 \pm 0.6M_\odot$ and radius $R=11^{+3}_{-2}$ km
for the compact star in 4U 1820-30.
Mass and radius of the compact star 
in 4U 1820-30 obtained in this paper are close to the canonical mass 
$M=1.4 M_\odot$ and radius $R=10$ km of neutron star (see the monograph 
by Haensel, Potekhin \& Yakovlev (2007)).

Our mass and radius determination are similar to the values published in
Shaposhnikov \& Titarchuk (2004), $R=11.2^{+0.4}_{-0.5}$ km and 
$M=1.29^{+0.19}_{-0.07}M_\odot$, respectively. Their estimates were based
on a simple analytical model of X-ray spectrum formation on the neutron 
star. Authors also assumed that the distance to the burster equals to 
$d=5.8$ kpc. However, our results differ from the conclusion obtained 
from analysis of quasi-periodic oscillations (QPO) in this source. It is
thought that the highest observed QPO frequency from 4U 1820-30 is the 
marginally stable orbit frequency, therefore, it implied the estimated 
mass of $M \sim 2.2 M_\odot$ (Smale, Zhang \& White 1997).

G\"uver et al. (2010) has just presented mass and radius estimation
for neutron star in 4U 1820-30. They obtained the most probable values 
$M=1.58 \pm 0.06 M_\odot$ and $R=9.1 \pm 0.4$ km from five strong bursts 
of this source observed by RXTE. Results of our paper are consistent
with that of G\"uver et al. (2010), though both papers presented different
approaches for fitting of the observed spectra.

In their research G\"uver et al. (2010) fitted observed X-ray spectra of the
source to the blackbody spectrum, corrected for interstellar absorption.
They subsequently determined other parameters, like normalisaton, bolometric
flux and blackbody temperature. Mass and radius could be eventually obtained
from set algebraic equations involving distance to 4U 1820-30 and the color
correction factor $f_c$, the latter was known from model atmosphere
calculations.

Our approach lies in the fact that shapes of both observed and numerical
spectra differ from shape of the blackbody. Then we performed trial fits of
a single observed X-ray count spectrum to hundreds of precomputed numerical
spectra (each of them corrected for interstellar absorption $N_H$), to find
the best fit $T_{\rm eff}$ and $\log g$, redshifted by $z$. No informations
on the distance, bolometric flux or color factors were applied here.

We note here that one cannot exclude, however, the possibility that the
momentary radius of the photosphere $R_{\rm ph}$ at touchdown phase
is still larger than the 'reference' radius $R$ of the neutron star in 
4U 1820-30 (Steiner et al. 2010, sec. 2.2).
In such a case our transformation from (z, log g) to M and R is not strictly
valid, cf. Eqs. 3-4. Then, errors of our M and R estimation must increase
substantially, as compared e.g. with those drawn in Fig. 3.

\section*{Acknowledgments}

This research was supported by the Polish Ministry of Science and Higher
Education grant No. N N203 511638.

\end{document}